\documentstyle[twoside,epsfig,fleqn,espcrc2]{article}

\newcommand{\PR}[3]{{ Phys. Rev.}        {\bf #1} {(19#2)} {#3}}

\newcommand{\AmS}{{\protect\the\textfont2
  A\kern-.1667em\lower.5ex\hbox{M}\kern-.125emS}}

\title{Models for Photon-photon Total Cross-Sections\thanks{
Supported in part by EEC-CT98-0169 and CICYT(AEN 96-1672)}}
\author{R. M. Godbole
\address{Centre for Theoretical Studies, Indian Institute of Science,
Bangalore, 560 012, India},
A. Grau\address{Departamento de F\'\i sica Te\'orica y del Cosmos,
Universidad de Granada, Granada, Spain}
and G. Pancheri\address{INFN-Laboratori Nazionali di
 Frascati, P.O.Box 13, I00044 Frascati, Italy}
}

\begin{document}
\begin{flushright}
IISc-CTS/7/99\\
UG-FT-102/99\\ 
LNF-99/019(P)\\
hep-ph/9908220\\
\end{flushright}
\begin{center}

{\Large Models for Photon-photon Total Cross-Sections}
\vskip 25 pt
{\large Rohini M. Godbole }\\
Centre for Theoretical Studies, Indian Institute of Science, Bangalore, 560 012, India.\\
\vskip 25 pt
{\large A. Grau}, \\
Departamento de F\'\i sica Te\'orica y del Cosmos, Universidad de Granada, 
Granada, Spain.\\
\vskip 25 pt
{\large G. Pancheri},\\
INFN-Laboratori Nazionali di
 Frascati, P.O.Box 13, I00044 Frascati, Italy\\

\vskip 50 pt
{\bf Abstract}
\\
We present here a brief overview of  recent models describing the
photon-photon cross-section into hadrons. We shall show in detail 
results from the eikonal minijet model, with and 
without soft gluon summation.
\end{center} 
\begin{abstract}
We present here a brief overview of  recent models describing the
photon-photon cross-section into hadrons. We shall show in detail 
results from the eikonal minijet model, with and 
without soft gluon summation.
\end{abstract}
\maketitle

\section{INTRODUCTION}

Due to the availability of  new  data from LEP on the total photon-photon 
cross-section\cite{L31,L32,OPAL1,OPAL2} in the  energy range $
\sqrt{s}\approx 10\div 150\ GeV$, we can now study a complete set of 
processes, in a similar 
energy region. This region covers
the  part where all total cross-sections are seen to rise. Thus one 
can now compare models, and their predictions, for proton-proton and 
proton-antiproton to those for photo-production 
as well as   photon-photon. In addition, the
increasing quantity of data from virtual photon processes allows for unique 
tests of our understanding of the role played by 
perturbative QCD in the rise of the cross-sections. Models for the 
the total $\gamma \gamma$ cross-section can be divided into three groups:
i) proton- like models, ii) LO QCD models, and iii) NLO QCD models. 
To the first category, 
one can ascribe the Regge-Pomeron type models, where the initial 
decrease and the subsequent rise are respectively attributed to the exchange 
of Regge and Pomeron trajectories, factorization models where a simple
constant allows to move from one process to the other, and scaling models 
where various inputs are scaled using Vector Meson Dominance (VMD) and 
Quark Parton Models (QPM).
QCD models ascribe the rise of the total cross-sections to LO QCD 
parton-parton scattering and NLO refine such models using higher order 
QCD effects, soft gluon summation, and/or $k_t$ parton distributions.

\section{MODELS WHERE THE PHOTON IS LIKE A PROTON}

In Fig.1 we show comparison between the latest data and curves from 
various models which treat the photon as an entity similar to
a hadron. In the Regge-Pomeron exchange model, the total cross-section is 
obtained from
\begin{equation}
\label{pom}
\sigma^{tot}=Y_{ab}s^{-\eta}+X_{ab}s^\epsilon
\end{equation}
with the power exponent taken to be the same for all processes while the
 coefficients obey the factorization-at-the-residue rule, i.e.
\begin{equation}
\label{factoriz} 
X_{aa}X_{bb}=X^2_{ab}\ \ \ \ \ \ \ Y_{aa}Y_{bb}=Y^2_{ab}
\end{equation}
The heavy full line shown in Fig. 1 is obtained using 
eqs.(\ref{pom},\ref{factoriz})
with  average values for the
Regge/Pomeron type description of all total cross-section,
i.e. $\eta=0.467$ and $\epsilon=0.079$\cite{PDG}.
This curve is 
not very different from a similarly inspired prediction by Schuler and 
Sj\"ostrand \cite{SAS1}. It interpolates between the two
data sets presented by L3~\cite{L32} using two different Montecarlo simulations,
Phojet and Pythia, and, at the same time, is a good description of the 
recently published OPAL data, averaged over Pythia and Phojet ~\cite{OPAL2}.
However, 
while the Regge/Pomeron model seems to
correctly predict the rise, it does not 
completely follow the trend of
the data, some of  which seem to rise faster\cite{L32} than with the power $\epsilon \approx 0.08$ 
extracted from the hadron-like processes. Notice however that, while this 
value of $\epsilon$ is a good fit to the
average rise in proton-antiproton \cite{DL}, according to the CDF 
Collaboration \cite{giromini} the power with which the proton-antiproton 
cross-section rises is $\epsilon=0.112\pm 0.0013$ \cite{giromini},
 a value closer to the power exponents implied by recent data on 
photoproduction obtained by  extrapolations from
Deep Inelastic Scattering data \cite{bernd} or the L3 data\cite{L31}.

\begin{figure}
\mbox{\epsfig{file=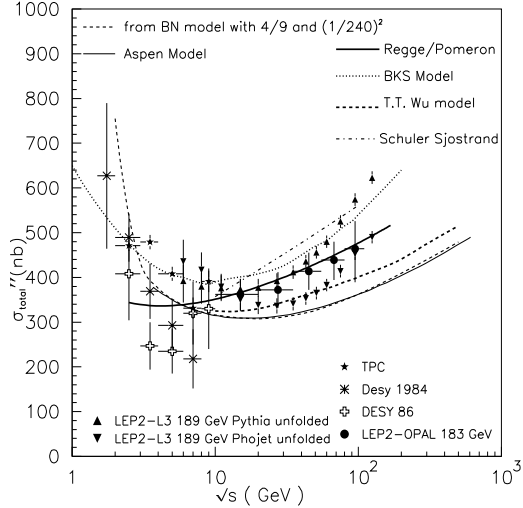,height=60mm,width=80mm}}
\caption{Data~\protect\cite{L32,OPAL2}  and 
different proton-like models for $\gamma \gamma$. 
Schuler and Sj\"ostrand  is from \protect\cite{SAS1}, 
the dash dotted curve from T.T. Wu et al. \protect\cite{ttwu},
Aspen model is from ref.\protect\cite{aspen} and BKS from 
\protect\cite{badelek}.}
\label{fig:largenenough}
\vspace{-0.5cm}
\end{figure}
Other models invoke simple scaling of the total cross-sections, i.e.
\begin{equation}
\sigma_{\gamma \gamma}=A \sigma_{pp}
\end{equation}
with the constant obtained either through  
factorization of the cross-section at high energy\cite{terazawa,ttwu}
 $A=({{\sigma_{\gamma p}}\over{\sigma_{pp}}})^2$
or using VMD and QPM to get $A=(2/3)^2 (1/240)^2$
where the first of these factors comes from quark counting and second
from the 
Vector Meson Dominance factor 
\begin{equation}
P_{had}=\sum_{V=\rho,\omega,\phi}{{4\pi\alpha}\over{f^2_V}}
\approx {{1}\over{240}}
\end{equation}
with
\begin{equation}
f_\rho=5.64, \ \ {{f_\rho}\over{f_\omega}}={{1}\over{3}}, \ \ 
{{f_\rho}\over{f_\phi}}={{-\sqrt{2}}\over{3}} 
\end{equation}
and 
$\alpha$ evaluated at the $M_Z$ scale.  The two very similar lowest curves in
 the figure are respectively obtained by a simple scaling (dashes) of a recent fit to 
proton proton, called BN model  and described in Sect.5, and
from the so called Aspen model, full curve\cite{aspen}. 
Both models are based on the eikonal approach. In
 the Aspen model \cite{aspen} 
\begin{equation}
\label{eik}
\sigma_{\gamma \gamma}=2 (P_{had})^2 \int d^2{\vec b}
[1-e^{-\chi_I(b,s)}cos{\chi_R}]
\end{equation}
and 
$\chi_I(b,s)=\sum W_{ij}(b,\mu_{ij})\sigma_{ij}(s)$, where the sum is over all 
possible parton type pairs, $\sigma_{ij}$ are elementary parton-parton 
cross-sections, $W_{ij}$ is an impact parameter distribution function, 
which is the Fourier-transform of the convolution of two dipole-type form 
factors, with scale factor $\mu_{ij}$. In
 this model the eikonal function  for $\gamma \gamma$ is obtained 
after fitting the proton-proton and
proton-antiproton cross-sections, 
by 
scaling of the s-dependence in the elementary cross-sections, i.e. 
$\sigma_{ij}^{\gamma \gamma}=(2/3)^2 \sigma_{ij}^{proton\ proton}$, 
and in the b-shape, i.e. $(\mu_{ij}^{\gamma \gamma})=3/2
 (\mu_{ij}^{proton \ proton})$. 
The curve labelled BKS model is extracted from  \cite{badelek} and is 
obtained 
from $F_2^{\gamma}$ in photoproduction, through a decomposition into
a VMD contribution and a QCD improved parton model term. Such
representation of the photon structure function is based on a similar 
representations of the
nucleon structure functions and as such can be included in the
 ``photon like a proton" types.

As mentioned in the introduction, these models do not introduce any 
different spatial structure of the photon, and obtain the total 
{
cross-section through extrapolation of some of the proton properties.
Finally, the highest of the  curves shown in Fig.1 is also from 
ref.\cite{SAS1}
and it corresponds to taking into account the `anomalous'
contributions to the photon structure, estimated using  $\gamma p$
data. 

\section{TOTAL CROSS-SECTIONS AND QCD}

These approaches have the ambitious aim to calculate
$\sigma_{tot}$ for any kind of process, through the same universal process 
independent  tools which are used in QCD for jet physics, namely 
parton-parton
subprocesses, parton densities and running $\alpha_s(Q^2)$. In these models
\cite{pythia,rohini,halzen,durand}
it is the rise with energy of the jet cross-section $\sigma_{jet}=
\int d^2{\vec p}_t{{d^2\sigma{jet}}\over{d^2{\vec p}_t}}$ which drives the
rise of the total cross-section. Since the early mini-jet models (the name comes
from the low  value of the jet $p_t$ needed to reproduce the high energy rise) 
do not satisfy unitarity in their simplest formulation, 
one has to use an improved 
theoretical framework\cite{durand} in which QCD processes can be embedded, like the eikonal 
formulation of eq.(\ref{eik}) in the approximation $\chi_R=0$ and 
$\chi_I=n(b,s)/2$, with $n(b,s)$ given by the average number of collisions at 
impact parameter $b$. In the Eikonal Minijet Model (EMM), the
average number $n(b,s)$ is schematically divided into a soft and a hard
component, i.e.
\begin{equation}
n(b,s)=n_{soft}(b,s)+n_{hard}(b,s)
\end{equation}
with the soft term basically determined by the bulk part of 
the inelastic cross-section, whereas it is left to $n_{hard}$ to drive the 
rise. Factorization of the impact parameter and energy dependence is almost 
certainly a good approximation
for what concerns the soft part of the interaction and one writes
\begin{equation}
n_{soft}(b,s)=A_{FF}(b)\sigma_{soft}(s)
\end{equation} with the b-dependence described through  convolution of the 
electromagnetic form factors of the colliding particles, namely a dipole form 
for the proton and a monopole form  for the photon, 
which is treated like  a meson or a pion for this purpose,
i.e.
\begin{equation}
A_{FF}(b)={{1}\over{(2\pi)^2}}\int d^2\vec{q}{\cal F}_a(q) {\cal F}_b(q) 
e^{i\vec{q}\cdot \vec{b}}
\end{equation}
This is often referred to as the Form Factor model (FF) for particles $a$ and $b$.
The soft s-dependence is modelled in different fashions,
 like a decreasing polynomial in $s$, with a Regge-type input
 with a 
scaling factor from hadron processes\cite{rohini}, etc. 
This is clearly the weakest part of all these models, but a part which can
be accessed perhaps later, after the high energy behaviour is 
well understood in QCD terms. For the high energy
piece of $n(b,s)$ there is still a large amount of
modelling, of course, mostly because of the b-dependence. A useful exercise 
is to start with
a b-dependence identical to the one used for the soft part, i.e, the
 Form Factor Model, and then adjust it after the s-dependence is fully
 understood. One such adjustment for the case of
proton-proton and proton-antiproton cross-section is described
in \cite{bn} where the overlap function in b-space is obtained through soft 
gluon summation of NLO terms in QCD. In all these models,
the main energy dependence, the actually observed rise of all total 
cross-sections, is
calculable and calculated from the QCD jet cross-section, integrated 
with $p_t\ge 1.5 \div 2\ GeV$, with currently used  proton and photon 
densities, and with running
$\alpha_s$. The procedure we adopt, to obtain the total photon-photon 
cross-section in the EMM, consists in first fixing  the arbitrary
low energy parameters in
photoproduction and then extrapolate them  to $\gamma \gamma$. The quantities 
to be
determined from photoproduction are the soft cross-section, the
parametrization of  the $b$-dependence,  the minimum $p_t$ 
over which to integrate the jet cross-section and the densities for the
partons in the photons. In Fig. 2 we show the $\gamma p$ cross-section data 
in comparison with three different parametrization  of the parton densities in
the photon: GRV \cite{GRV}, SAS\cite{SAS} and GRS \cite{GRS}, 
for different choices of $p_{tmin}$. 
\begin{figure}[htb]
\mbox{\epsfig{file=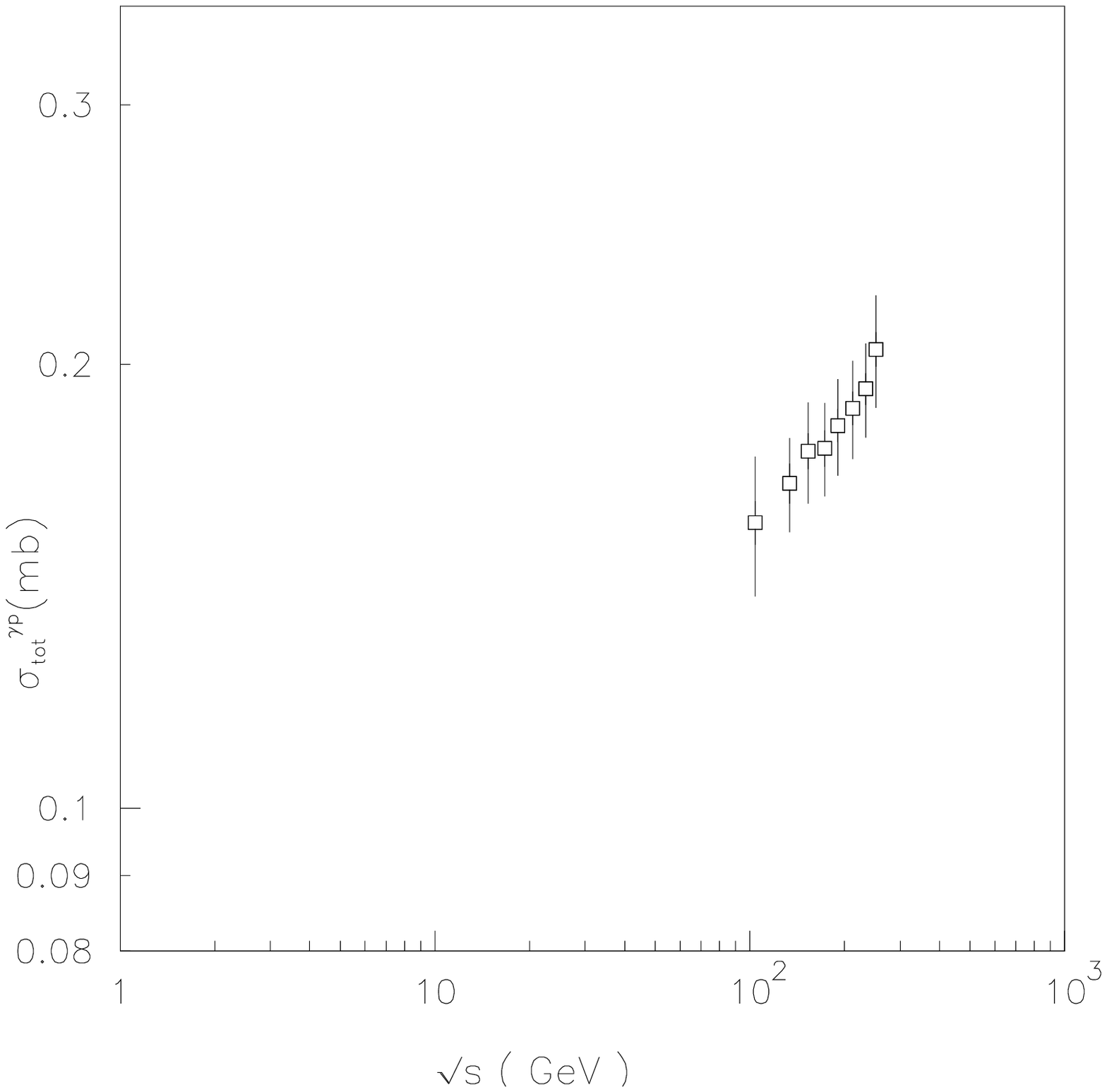,height=60mm,width=80mm}}
\caption{Comparison of photoproduction and extrapolated DIS data with EMM 
model}
\label{fig:gamp}
\vspace{-0.5cm}
\end{figure}
The two sets of data correspond
 respectively to photoproduction \cite{ZEUS,H1} and to a recent 
extrapolation of the `nonzero-$Q^2$' data \cite{bernd}.  In all these the 
parametrisation  for the parton densities in the proton is GRV94~\cite{grvpr}. 
We have checked that the results are not affected if we use the 
GRV98~\cite{grvpr1} parametrisation instead.  As said before, 
the $b$ dependence can also be modelled by Fourier 
Transform of  the intrinsic transverse  momentum 
distribution of the  partons. For the photonic partons this gives rise to a
$b$ dependence  which  has the same analytical form as the form factor ansatz,
but with a different value for the pole position~\cite{rohini} than with the
form factor ansatz. All the curves correspond to using for the photon 
``form factor" 
\begin{equation}
{\cal F}(q^2)={{k_0^2}\over{q^2+k_0^2}}
\end{equation}
with $k_0=0.66\ GeV$ a value implied by the measurements by ZEUS, 
a value of $P_{had}=1/240$ and a soft cross-section 
$\sigma_{\gamma p}=constant +A/\sqrt{s}+B/s$.

%
With these values of the various parameters, one can obtain the photon-photon 
cross-sections. To do so, we use
$P_{had}^{\gamma \gamma}=[P_{had}]^2$, 
$\sigma_{soft}^{\gamma \gamma}=2/3\sigma_{soft}^{\gamma p}$ and the
same values for the $k_0$ scale in the photon form factor, same densities and $p_{tmin}$ values.
The result is shown in Fig.3.

\begin{figure}[htb]
\mbox{\epsfig{file=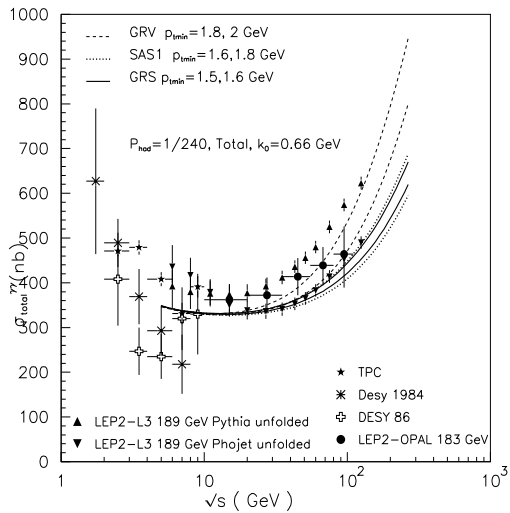,height=60mm,width=80mm}}
\caption{Comparison of photon-photon total cross-section data with EMM 
model}
\label{fig:gamgam-emm}
\vspace{-0.5cm}
\end{figure}

\section{INELASTIC AND TOTAL PHOTON CROSS-SECTIONS IN THE EMM}

We  compare here  two different applications of the eikonal 
minijet model
 for
the photon cross-sections into hadrons, the {\it total} vs. {\it inelastic} mode.
Namely,  curves  are
obtained by fixing the parameters as
described in the previous section, but using a fit to the
photoproduction data considered as an {\it inelastic} cross-section. 
This kind of comparison would correspond to a situation such that the
data did not include any of the so-called elastic part, namely vector
meson production. The latter evaluation depending upon
the model used for diffractive production, fitting the curve with 
\begin{equation}
\label{eikinel}
\sigma^{inel}=P_{had} \int d^2{\vec b}[1-e^{-n(b,s)}]
\end{equation}
and then extrapolating to $\gamma \gamma$, again
with the inelastic formulation, 
may provide an interpolation for the actual data for photon-photon
between different models for the diffractive part.
Proceeding as described, we obtain a slightly different set of parameters 
than the ones described
in the previous section, including a value for $P_{had}\approx 1/200$ and
a curve for photon -photon which lies higher than the previous one, with 
a more moderate rise. 
\begin{figure}[htb]
\mbox{\epsfig{file=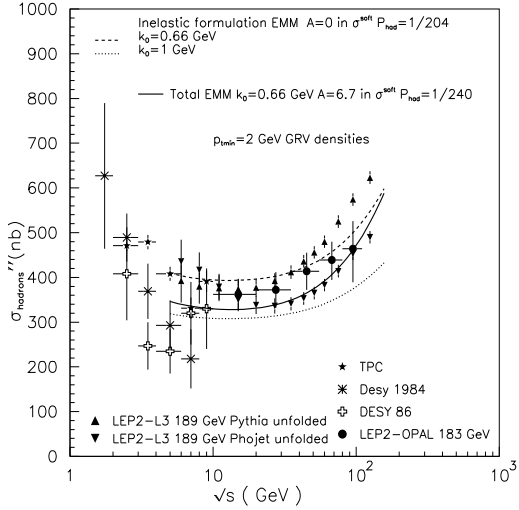,height=60mm,width=80mm}}
\caption{Comparison of photon-photon total and inelastic cross-section 
predictions from the EMM.}
\label{fig:gamgam_tot}
\vspace{-0.5cm}
\end{figure}
The results are shown in Fig.~\ref{fig:gamgam_tot}. As
first noticed in \cite{aspen}, the total EMM prediction is more in 
agreement with the L3 data 
(as extrapolated with Phojet),
whereas the inelastic formulation follows better the trend of the OPAL data.
 To illustrate the dependence of the EMM  predictions upon the 
scale parameter $k_0$, 
the inelastic formulation is shown with two curves, the lower of which is 
obtained with  $k_0=1\ GeV$ value.

\section{QCD MODEL FOR THE OVERLAP FUNCTION A(b)}
In this last section we present some work in progress concerning a soft 
gluon summation model for the overlap function
$A(b)$, which should hopefully reduce some of the arbitrariness of the
mini-jet model, namely the dependence upon the impact parameter space 
distribution. In the previous section, the average number of hard collisions 
was given as
\begin{equation}
n_{hard}(b,s)=A(b)\sigma_{jet}(s,p_{tmin})/P_{had}
\end{equation}
where $A(b)$ can be considered as the average number of scattering centers per 
unit area and, as mentioned, is obtained by  convoluting the
electromagnetic form factor of particles $a$ and $b$. The philosophy 
underlying this assumption is 
that not only matter distribution in the hadron follows the charge distribution,
but also the quantochromodynamic field, the gluons, does. While this is 
certainly plausible, it does not allow for an {\it ab-initio}
description. The parton-model
improved ansatze of \cite{bn} is that there is no exact factorization between the
variables in the transverse and longitudinal momentum, and that each 
subprocess characterized by a final jet momentum
$p_t$ be defined through an impact parameter value $b$ whose distribution is
obtained as the
Fourier transform of the initial transverse momentum of the colliding pair,
 which was
initially assumed collinear with the proton.
In this model, which includes NLO corrections to the leading order Born 
parton-parton processes, one has
\begin{equation}
A(b)={ {e^{-h(b,s)}}
\over{\int d^2{\vec b}\  
e^{-h(b,s)}
}}
\end{equation}
where 
\begin{equation}
e^{-h(b,s)}=\int {{d^2 {\vec K}_t}\over{(2\pi)^2}} \Pi(K_t)e^{-iK_t\cdot b}
\end{equation} and $\Pi(K_t)$ is the transverse momentum distribution of 
initial state soft gluon emission in the process
\begin{equation}
 q {\bar q} \rightarrow X + {\rm jet\;\; jet}
\end{equation}
In LLA, the function $h(b,s)$ is calculated from QCD to be given by\cite{bn}
\begin{eqnarray}
\label{hdb}
h(b,s)={{2 c_F}\over{\pi}}\int_0^M {{dk_\perp}
\over{k_\perp}}\alpha_s({{k^2_\perp}
\over{\Lambda^2}}) \nonumber \\
\ \ \ \times \ln{{M+\sqrt{M^2-k_\perp^2}}\over{M-\sqrt{M^2-k_\perp^2}}}
[1-J_0(k_\perp b)]
\end{eqnarray}
where $M\simeq M(s,p_{tmin})$ is the maximum energy allowed for gluon emission 
in a process
characterized by a given $p_{tmin}$ at any given energy $\sqrt{s}$.
Thus the Bloch-Nordsieck improved b-distribution function is 
energy dependent, since the function $h(b,s)$ depends upon
the maximum energy allowed for single gluon emission by each colliding parton,
and  a rather lengthy and complicate integration may be involved.
\begin{figure}[htb]
\mbox{\epsfig{file=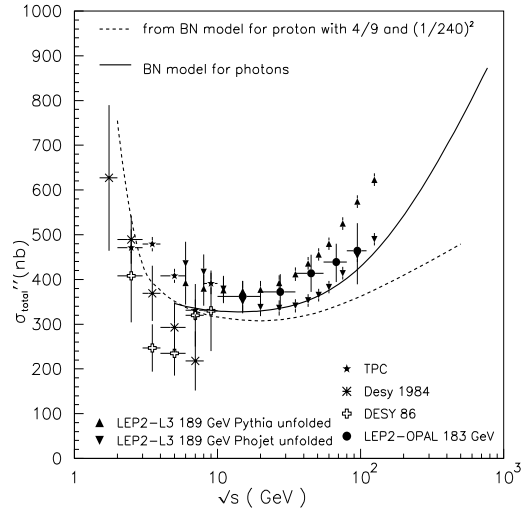,height=60mm,width=80mm}}
\caption{Predictions from the EMM implemented by Bloch-Nordieck summation.} 
\label{fig:gamgam_inel}
\vspace{-0.5cm}
\end{figure}
Such impact parameter distribution inserted into the eikonal formalism of the
minijet model describes a picture of multiparton collisions, each one of 
which is dressed by
soft gluons. While this model has been recently\cite{bn} 
shown to give interesting 
results for the
case of proton-proton, and proton-antiproton, work
is in progress to determine the effect of such distribution on photon-photon 
cross-sections. A preliminary result is shown in Fig.5.

\section{Conclusion}
We have presented various predictions for the total $\gamma \gamma$ 
cross-section into hadrons,
some of them based on a straightforward extension of the proton 
total cross-sections, others
which depend more strongly on the chosen parton densities in the
photons at small x. We notice that the
rise of the total photon-photon cross-section can be accounted for
through the QCD mini-jet models.
We also see that more work is still needed, both
 on the theoretical and the
experimental side, before being able to make realistic descriptions.

\section{Acknowledgments}
We wish to thank Prof. Martin Block for useful discussions.

\end{document}